\documentclass[]{raa}
\usepackage{amssymb}
\usepackage{graphicx,times}
\usepackage{natbib}

\providecommand{\bm}[1]{\mbox{\boldmath$#1$\unboldmath}}

\begin{document}

   \title{Why Are Halo Coronal Mass Ejections Faster?}

 \volnopage{ {\bf 2010} Vol.\ {\bf X} No. {\bf XX}, 000--000}
   \setcounter{page}{1}

   \author{Q. M. Zhang$^1$
   \and Y. Guo$^1$
   \and P. F. Chen$^{1,2}$
   \and M. D. Ding$^{1,2}$
   \and C. Fang$^{1,2}$
   }

   \institute{$^1$ Department of Astronomy, Nanjing University, Nanjing 210093,
    China; {\it chenpf@nju.edu.cn}\\
	$^2$ Key Lab of Modern Astron. and Astrophys., Ministry of Education,
	China\\
\vs \no
   {\small Received [year] [month] [day]; accepted [year] [month] [day]}
}

\abstract{Halo coronal mass ejections (CMEs) were found to be
significantly faster than normal CMEs, which was a long-standing puzzle.
In order to solve the puzzle, we first investigate the observed
properties of 31 limb CMEs that display clearly loop-shaped frontal
loops. The observational results show a strong tendency that slower CMEs
are weaker in the white-light intensity. Then, we perform a Monte Carlo
simulation of 20\,000 artificial limb CMEs that have average velocity of
$\sim$523 km s$^{-1}$. The Thomson scattering of these events is
calculated when they are assumed to be observed as limb and halo
events, respectively. It is found that the white-light intensity of many
slow CMEs becomes remarkably reduced as they turn from being viewed as
a limb event to as a halo event. When the intensity is below the
background solar wind fluctuation, it is assumed that they would be
missed by coronagraphs. The average velocity of ``detectable'' halo CMEs
is $\sim$922 km s$^{-1}$, very close to the observed value. It also
indicates that wider events are more likely to be recorded. The results
soundly suggest that the higher average velocity of halo CMEs is due to
that a majority of slow events and a part of narrow fast events carrying
less material are so faint that they are blended with the solar wind
fluctuations, and therefore are not observed.
    \keywords{Sun: coronal mass ejections (CMEs)
 --- Sun: activity --- methods: numerical --- solar-terrestrial relations} }

   \authorrunning{Q. M. Zhang et al.}
   \titlerunning{Why Are Halo CMEs Faster?}
   \maketitle

\section{Introduction} \label{intro}

Since the first recognition of coronal mass ejections (CMEs,
initially called coronal transients) by OSO-7 \citep{tous73}, more
than 10\,000 such energetic events have been identified by
ground-based and space-borne coronagraphs. Although remarkable
progress had been made before, the Large Angle Spectrometric
Coronagraph (LASCO) \citep{brue95}, which is aboard the Solar and
Heliospheric Observatory (SOHO) spacecraft \citep{domi95} launched
at the end of 1995, revolutionized our understanding of this
eruptive activity for its large field of view (FOV), increased
sensitivity and dynamic range. CMEs are often associated with solar
flares and filament eruptions \citep{chenaq06, chenaq09}, leading to
large-scale coronal disturbances like EIT waves \citep{chenpf06,
chenpf09}, and even triggering a sympathetic CME \citep{cheng05a}. A
typical CME exhibits the 3-part morphology:
a frontal loop, which is followed by a dark cavity with an embedded
bright core \citep[e.g.,][]{Ill85, Dere97}. It expels approximately
$10^{14}$-$10^{16}$ gram \citep[e.g.,][]{webb95} of plasma into
interplanetary space with velocity ranging from tens to 3000 km s$^{-1}$
\citep[e.g.,][]{stcyr92, hund94}, and with kinetic energy up to 
$\sim$10$^{31}$ erg \citep{burk04}. At the same time, magnetic field and its
intrinsic magnetic helicity are ejected into interplanetary space,
which plays an essential part in completing the global magnetic
field reversal between successive solar cycles
\citep[e.g.,][]{zhang05}. They can potentially give rise to
hazardous terrestrial effects, such as solar energetic particles
\citep{reames99}, type II radio burst \citep{gopal09}, geomagnetic
storms \citep[e.g.,][]{gosl90}, ionosphere disturbance, and
polar aurorae. The information of the magnetic
field, mass, and velocity of CMEs is very important since they
determine the geomagnetic effectiveness \citep{gopal07}.

As a special type, those surrounding the occulting disk, i.e., with
an apparent angular width of $360^\circ$, are called full halo CMEs
\citep{howa82}. Compared with normal events with apparent
widths between $20^\circ$ and $120^\circ$, they are generally
believed to be nothing special except that they propagate in a
direction close to the Sun-Earth line, either toward or away from the
Earth. However, it has been noticed that the average apparent
velocity of halo CMEs is fairly higher than that of normal
CMEs \citep{webb99}. For example, \citet{yashiro04} compared
$\sim$7000 events in the period from 1996 to 2002, and
found that the average apparent velocity of halo CMEs is twice
larger than that of normal CMEs. Such a difference made
\citet{lara06} propose that halo CMEs are of a special type. Realizing
that white-light emission of CMEs comes mainly from the
Thomson-scattering of photospheric radiation, which is much
weaker for plasma near the Sun-Earth line than that near the plane
of the sky at the same projected heliocentric distance,
\citet{andr02} suggested that there exists a mass cut-off, above
which CMEs are bright enough to be detected, while many dim,
presumably slow events, are missed by coronagraphs. This
would make the average apparent velocity of halo CMEs much higher
than that of the normal type.

Such a conjecture can be validated if slower CMEs are systematically
fainter in brightness. For this purpose, \citet{cheng05b} studied the
relationship between the apparent velocity and the white-light intensity
for halo CMEs. The two parameters did show a marked positive
correlation, which provides an indirect support for the mass cut-off
conjecture.  Compared to limb events, halo CMEs should travel a longer
distance to be observed in the FOV of coronagraphs, which has two
effects making halo CMEs significantly fainter in white light. First,
the intensity of the incident emission from the photosphere is lower.
Second, the number density of the CME front becomes smaller. Despite
that the cross-section of the Thomson scattering of halo CMEs increases
\citep[see][] {bill66}, their white-light emission is reduced greatly
compared to the limb CMEs events that are observed at the same projected
distance in the plane-of-the-sky. Therefore, it is expected that many
halo CMEs, especially the slower events, could be so faint that are
missed by coronagraphs. In this paper, we collect a sample of 31 limb
CMEs free from projection effect during the SOHO Mission to confirm the
velocity-brightness relation. A Monte Carlo simulation is further
performed to quantitatively testify whether the observed high value of
the average velocity of halo CMEs is due to that many slow events are
missed by coronagraphs.

This paper is organized as follows. Data analysis and the results are
presented in Section~\ref{analysis}. The Monte Carlo simulation and its
result are shown in Section~\ref{mtkl}, followed by the discussion on
the projection effects in Section~\ref{proj}. We summarize the results
in Section~\ref{sum}, along with some discussions.

\section{Data Analysis and Results} \label{analysis}

Located at the inner Lagrangian point ({\it $L_1$}), SOHO has been
monitoring the vigorous Sun for 14 years. Three coronagraphs,
C1, C2, and C3, which constitute the LASCO instrument, have FOVs of
1.1-3$R_\odot$, 2-6$R_\odot$, and 4-32$R_\odot$, respectively, where
$R_\odot$ is the solar radius. The routine
observations provide a huge database for CME research.

\begin{table}
\small \centering

\begin{minipage}[]{120mm}
\caption[]{List of basic properties of the 31 loop-shaped limb CMEs,
including the date, time of first appearance in C2 FOV, central
position angle (CPA), angular width (AW), and the linear velocity
($V$). }\label{Table 1}
\end{minipage}

\tabcolsep 6mm
\begin{tabular}{ccccc}
\hline\noalign{\smallskip}
Date & Time & {CPA} & {AW} & {$V$}\\
     & (UT) & (Deg) & (Deg) & (km s$^{-1}$)\\
\hline\noalign{\smallskip}
1997/09/29 & 18:30 &  78 &  99 & 369\\
1997/12/05 & 08:27 & 275 &  98 & 414\\
1998/01/28 & 14:56 & 268 &  74 & 246\\
1998/02/25 & 23:27 &  74 &  65 & 289\\
1998/11/09 & 01:54 &  16 &  94 & 144\\
1999/05/17 & 00:50 & 293 & 113 & 503\\
2000/08/12 & 15:54 & 254 & 117 & 499\\
2000/08/22 & 23:06 & 179 &  59 & 431\\
2000/11/27 & 23:54 & 123 &  57 & 474\\
2001/01/26 & 16:06 &  55 & 111 & 698\\
2001/05/28 & 23:50 &  96 &  41 & 542\\
2001/06/13 & 00:06 & 279 &  62 & 447\\
2001/08/30 & 09:50 & 129 &  86 & 462\\
2001/09/05 & 16:06 & 232 & 107 & 538\\
2002/01/10 & 00:30 & 236 &  61 & 377\\
2002/03/12 & 23:54 & 112 &  82 & 535\\
2002/03/15 & 02:06 &  64 &  65 & 272\\
2002/04/18 & 06:26 & 162 &  64 & 552\\
2002/09/18 & 14:54 & 279 &  99 & 512\\
2002/11/04 & 12:30 &  17 & 114 & 509\\
2002/11/08 & 11:30 & 298 &  69 & 424\\
2003/01/03 & 11:30 & 283 &  88 & 521\\
2003/01/20 & 18:30 & 315 & 105 & 733\\
2003/03/21 & 10:54 &  54 &  66 & 481\\
2003/12/08 & 13:31 & 228 &  68 & 464\\
2004/05/03 & 00:50 & 113 & 112 & 464\\
2004/07/10 & 13:54 & 270 &  78 & 477\\
2004/08/18 & 17:54 & 258 & 120 & 602\\
2004/08/27 & 09:30 & 261 &  70 & 554\\
2005/03/14 & 08:00 & 259 & 105 & 849\\
2005/09/04 & 14:48 & 286 &  86 & 1\,179\\
\hline
\end{tabular}
\end{table}

In order to select limb events, we carefully checked the movies that
are composed of LASCO/C2 and EIT \citep{Dela95} difference images
from the NASA CME
catalog\footnote{http://cdaw.gsfc.nasa.gov/CME\_list} between
1997 January and 2005 December. The events with associated flares or
filament eruptions occurring beyond the longitude of $50^{\circ}$
are collected. Moreover, only the CMEs that have definitely loop-shaped
leading edge are considered. It should be kept in mind that the bright loop is
not a simple tube but the projection of the dense front of
three-dimensional bubble-like structure. To select events with clear loop tops,
those CME events with their legs being much brighter than the tops are
ignored.
Besides, some events whose white-light intensity increases
with height or can poorly be fitted with a power-law function with height
are also excluded (possibly due to that they are undergoing
acceleration). As a result, the sample finally consists of 31
well-defined events, whose image quality indices are $\geq$ 4 (labeled
as ``good'' events in the CME catalog). To our understanding, the events
with high quality appear less diffuse and have sharp contrast to the
background corona so that the
height-time measurement is more precise. The basic properties, including
the observation date, time of first appearance in the C2 FOV, central
position angle (CPA), angular width (AW), and the linear velocity ($V$),
are listed in Table~\ref{Table 1}.

\begin{figure}
   \centering
   \includegraphics[width=7.0cm, angle=0]{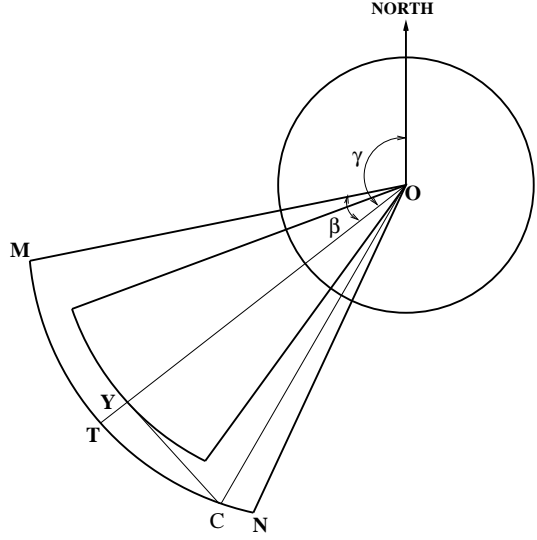}

   \caption{A sketch of the plane-of-the-sky-projected axisymmetric cone
    model applied to the analysis of loop-shaped limb CMEs at their early
    stage of propagation in the corona. The cone OMTNO outlines the CME
    leading edge, while $\gamma$ and $\beta$ are the central position
    angle and the angular half-width, respectively. Line {\it CY}
    perpendicular to {\it OT} equals to half LOS depth of $Y$.
   }
   \label{Fig1}
   \end{figure}

\begin{figure}
   \centering
   \includegraphics[width=7.0cm, angle=0]{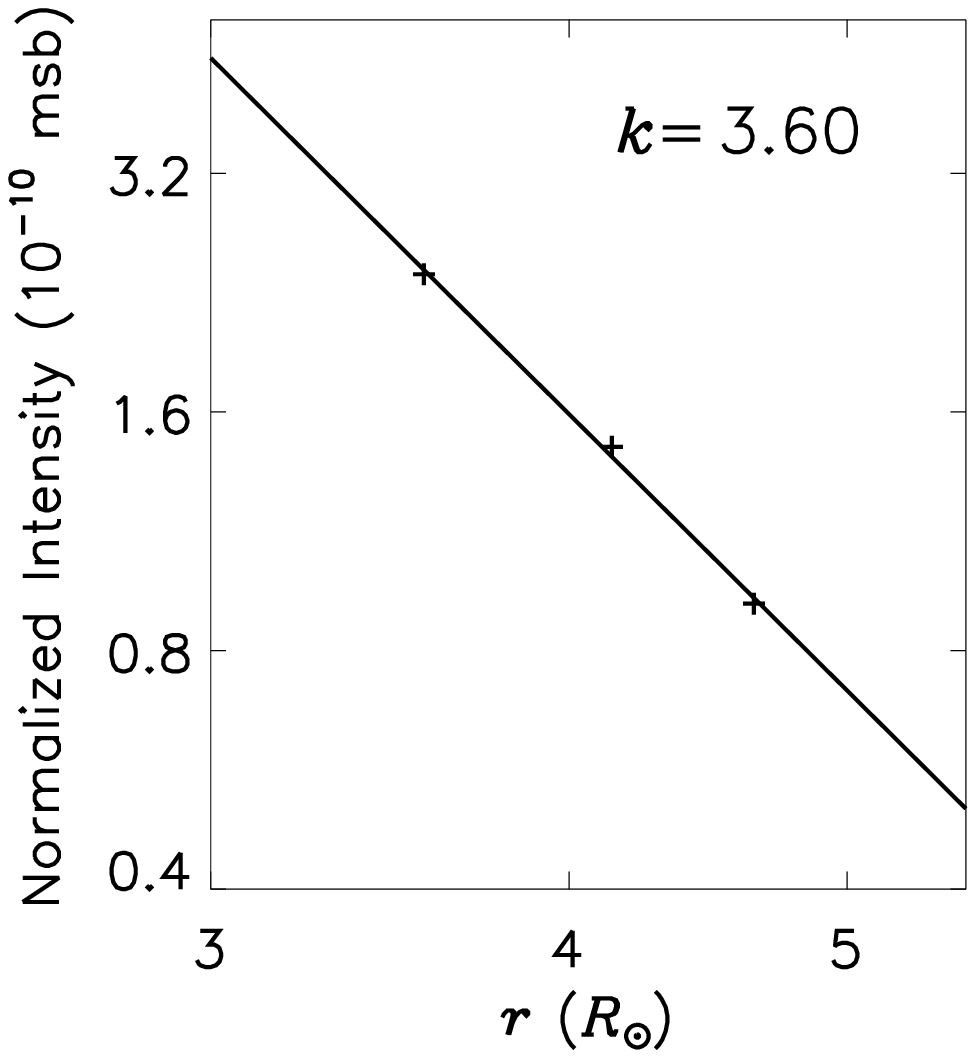}
   \caption{Variation of the normalized intensity $I_n$ of the CME leading
    edge with the heliocentric distance $r$ for the 2000 August 22 event.
    The variation is fitted with a power-law profile $I_{n}\sim r^{-k}$,
    where $k=3.60$.
   }
   \label{Fig2}
   \end{figure}

Pre-processing of the LASCO/C2 data, such as correction of flat
field and removal of dark current, is conducted by using the
standard program {\it c2\_calibrate.pro} in the Solar SoftWare
(SSW). For each event, we take the white-light image just before the
first appearance in the C2 FOV as base image to get base difference
intensity of the CME frontal loop in the ensuing three snapshots.

\begin{figure}
   \centering
   \includegraphics[width=7.0cm, angle=0]{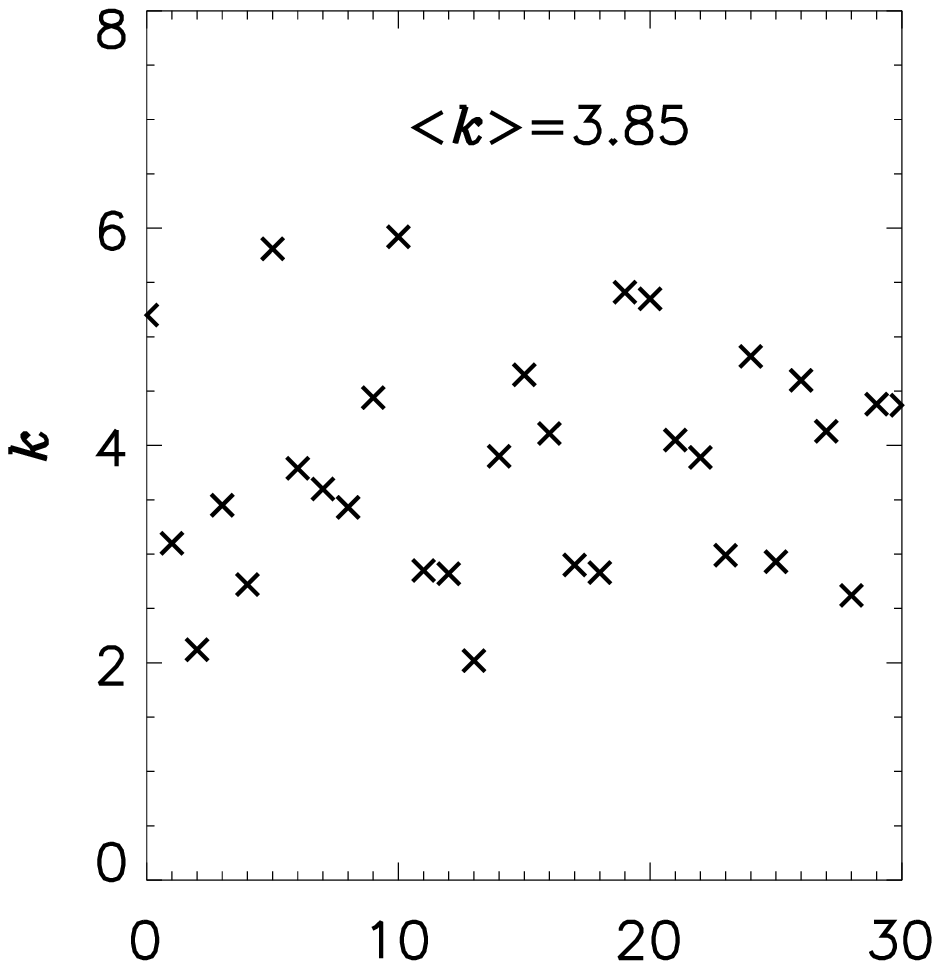}
   \caption{Scatter plot of the power index $k$ in the 31 loop-like
   events. The average value of $k$ is 3.85.
   }
   \label{Fig3}
   \end{figure}

It is noted that the observed white-light intensity is the emission
integrated along the line-of-sight (LOS), which represents the total
mass of unit area in the plane-of-the-sky, i.e., column density. We call
it integral intensity. In order to better characterize volume density
enhancement of the CME leading loop, we define another parameter --
normalized intensity, which is the integral intensity divided by the
LOS depth in unit of $R_\odot$. Since the 3D topology of CMEs is unclear
from a single viewing direction, several versions of the cone model have
been proposed based on the fact that the apparent angular widths of many
CMEs remain almost constant during their propagation in the corona
\citep[e.g.,][]{Anzer79, Fisher84, Mich03, Xie04}. According to
\citet{Schw05}, the projected geometry that can well
reproduce the kinematic properties of CMEs with the cone angle
between 40$^\circ$ and 80$^\circ$ is displayed in Fig.~\ref{Fig1}.
The frontal loop is concentric with the solar disk. The position
angle of cone axis {\it OT} is denoted by $\gamma$, and the angular
half-width by $\beta$, respectively. The LOS depth of white-light
emission, which is an unknown parameter, is supposed to be of the same
size as the transverse expansion according to the cone model.
In the right triangle $\bigtriangleup OYC$ ($Y$ denotes a
point along the line {\it OT} within the frontal loop), $OC=OT$ and
$YC=\sqrt{(OC)^2-(OY)^2}=\sqrt{(OT)^2-(OY)^2}$. Under the
axisymmetry assumption of the CME 3D morphology, the LOS depth at
$Y$ equals to 2{\it YC}.

\begin{figure}
   \centering
   \includegraphics[width=13.0cm, angle=0]{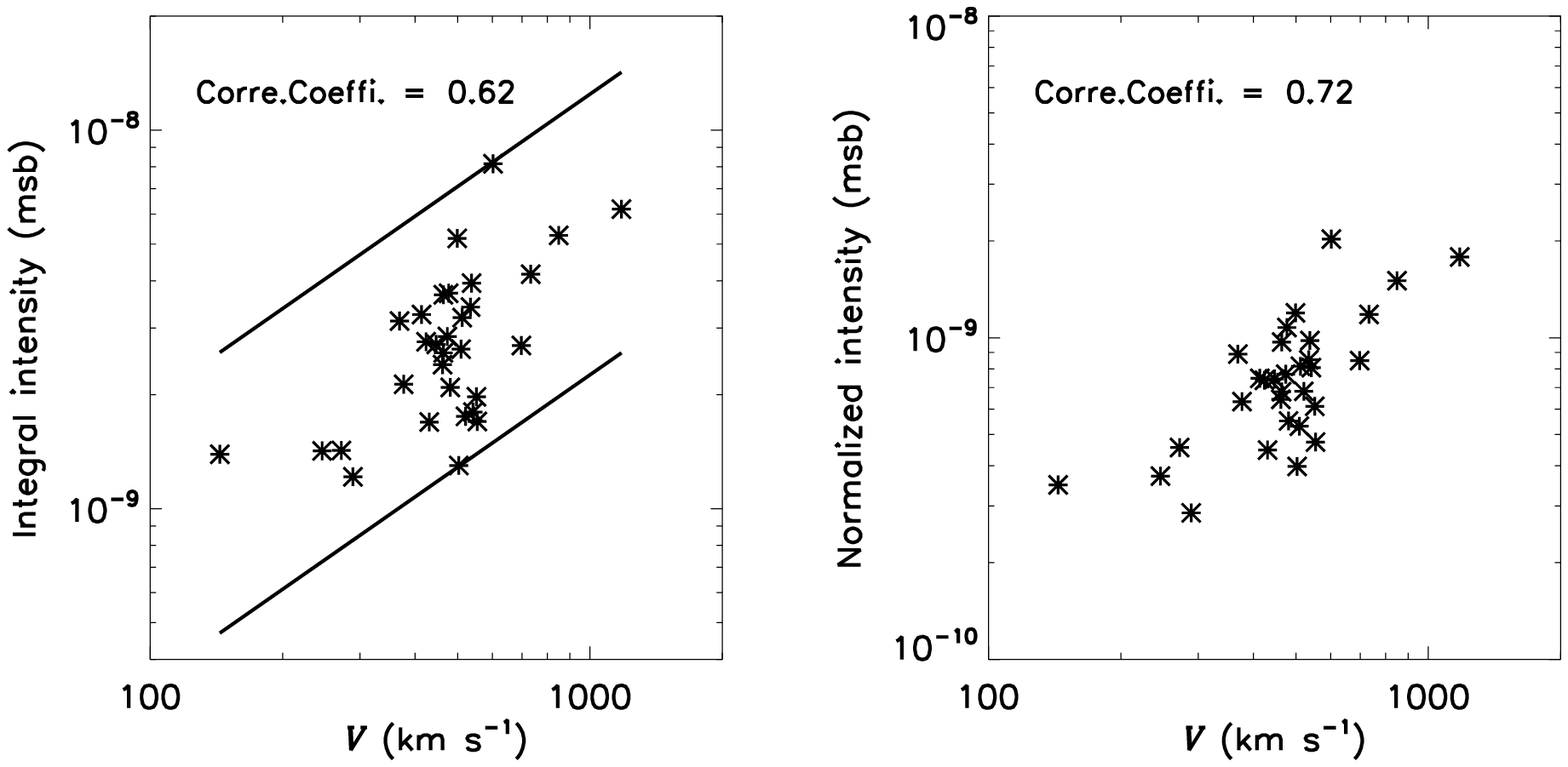}
   \caption{Scatter diagrams showing the relationship between the
     velocity and the integral ({\it Left panel}) and normalized
     ({\it Right panel}) intensity at the loop center of the 31
     sample events. The solid lines bordering the data points in the
     left panel are parallel to the fitted line from these points.
     The correlation coefficients are displayed at the upper left of
     the panels.
   }
   \label{Fig4}
   \end{figure}

For each CME, the normalized intensity of the loop-top at three moments
is obtained and fitted with a power-law function $I_{n}\sim r^{-k}$,
where $r$ and $k$ are heliocentric distance and the power index,
respectively. As an example, we plot the variation of the normalized
intensity $I_n$ with $r$ for the 2000 August 22 CME event in Fig.
\ref{Fig2}, which corresponds to $k=3.60$. In the same way, we derive
$k$ for each loop-like event based on near-perfect power-law regression.
The distribution of the index, which is averaged at 3.85, is shown in
Fig.~\ref{Fig3}, i.e., on avarage, $I_n$ decreases with the heliocentric
distance $r$ as $I_{n}\sim r^{-3.85}$. As mentioned in Section
\ref{intro}, two factors contribute to the decrease of $I_n$ with $r$.
One is the incident emission from the photosphere, the other is the
number density of the CME front. It is known that the incident emission
decreases with $r$ as $r^{-2}$, therefore, it is derived from the 31
limb events that the number density of the CME frontal loop decreases
with height as $\sim r^{-1.85}$, a little more slowly than the first
factor. Since a flare- or filament-related CME
usually undergoes three phases: initiation phase, fast acceleration
phase in the inner corona ($\leq$ 3$R_\odot$), and propagation phase
with constant speed \citep{zhang01,zhang04}, the white-light intensity
interpolated at 3$R_{\odot}$, both before and after normalization, is
calculated to check the relation between CME velocity and brightness.
For the 2000 August 22 event, the integral and normalized intensity at
3$R_{\odot}$ is $1.70\times 10^{-9}$ msb and $4.47\times 10^{-10}$
msb, respectively. Here, ``msb'' is short for ``mean solar brightness''.

The relation between CME velocity and integral intensity at the frontal
loop center (which is also the brightest point) is shown in the left
panel of Fig.~\ref{Fig4}. A clear tendency is seen that the integral
intensity of the CME leading loop increases with the CME velocity. The
correlation coefficient is as high as 0.62.  Apparently, the 31 limb
CMEs all fall in the domain bracketed by the two parallel solid lines in
the panel. The relation between CME velocity and the normalized
intensity at the frontal loop center is shown in the right panel of Fig.
\ref{Fig4}. The correlation coefficient increases to 0.72 after
normalization. Moreover, the scattering of the data points is reduced
compared with the left panel.

The positive correlation between the white-light intensity and the
velocity of the limb events confirms the result of \citet{cheng05b}, and
provides indirect evidence in favor of the conjecture of \citet{andr02},
who proposed that the high average halo CME velocity is due to that some
slow events are neglected by coronagraphs. To quantitatively justify
this viewpoint, we carry out a Monte Carlo simulation in the next
session on the basis of the above correlation.

\begin{figure}
   \centering
   \includegraphics[width=11.0cm, angle=0]{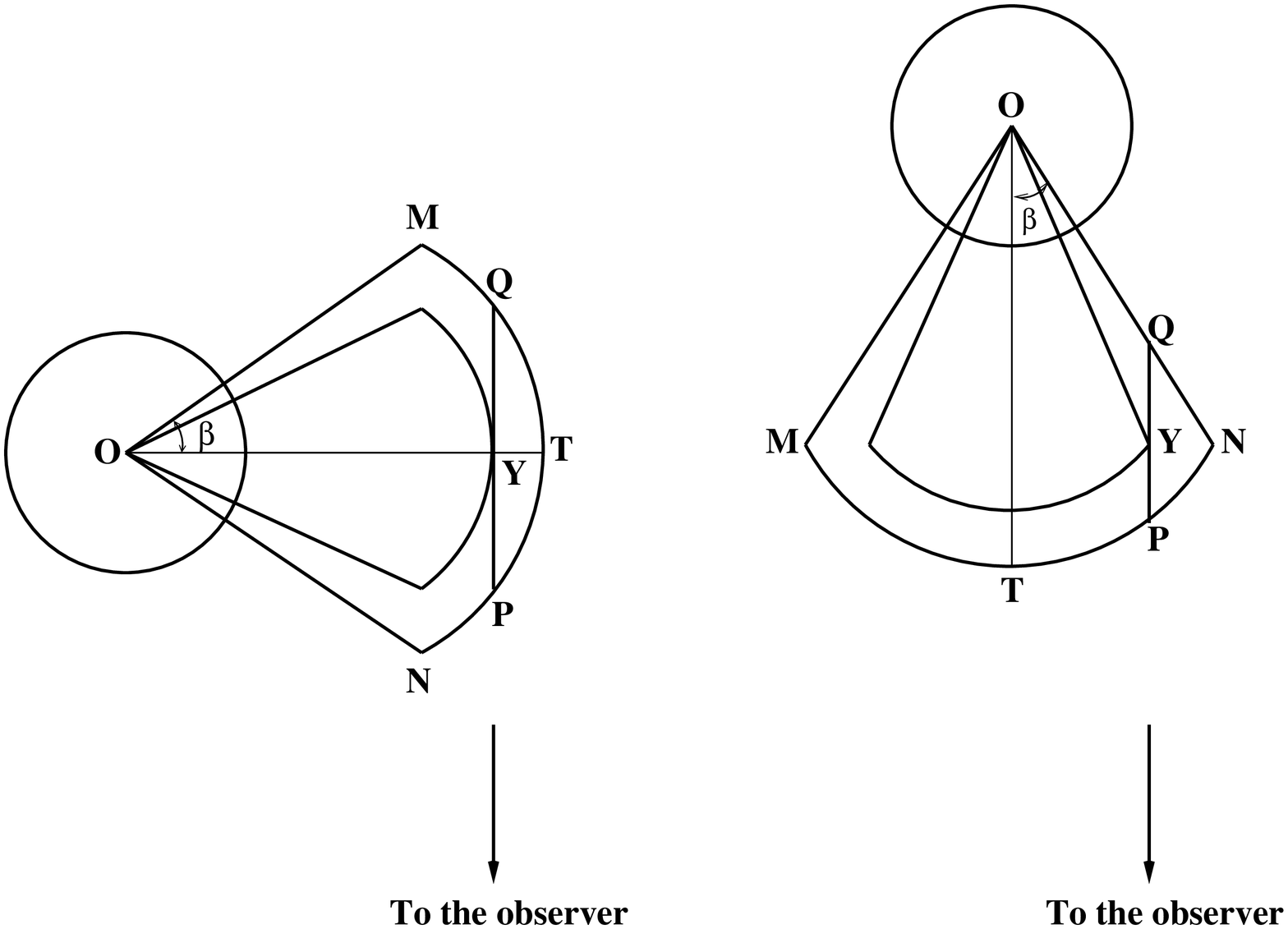}
   \caption{Two sketches illustrating how white-light coronagraphs view
CME from an edge-on ({\it Left panel}) and face-on ({\it Right
panel}) perspective.
   }
   \label{Fig5}
   \end{figure}

\section{Monte Carlo Simulation} \label{mtkl}

As mentioned in Section~\ref{intro}, the white-light emission scattered
by the CME front in the halo events would be greatly reduced, compared
to limb events. Once the white-light intensity falls below the
fluctuation level of the background solar wind, it is likely that a halo
CME event is missed by coronagraphs. In this section, using Monte Carlo
simulation of 20\,000 artificial CMEs, we attempt to quantitatively
estimate how these effects change the average velocity of the observed
halo CMEs.

Figure~\ref{Fig5} depicts the geometry of a limb ({\it left panel}) and
a halo ({\it right panel}) CME, based on which we now check how a limb
event changes in the apparent velocity and the white-light intensity
when it is observed as a halo event. From the geometry, it is easy to
see that the apparent velocity decreases by a factor of $\sin \beta$
when a limb CME is observed face-on as a halo event, where $\beta$ is
the true angular half-width. Line $PQ$ in each panel stands for the
longest LOS depth of the cone shell. The integral intensity in the limb
and halo cases are labeled with $I_L$ and $I_H$, respectively.

\begin{figure}
   \centering
   \includegraphics[width=7.0cm, angle=0]{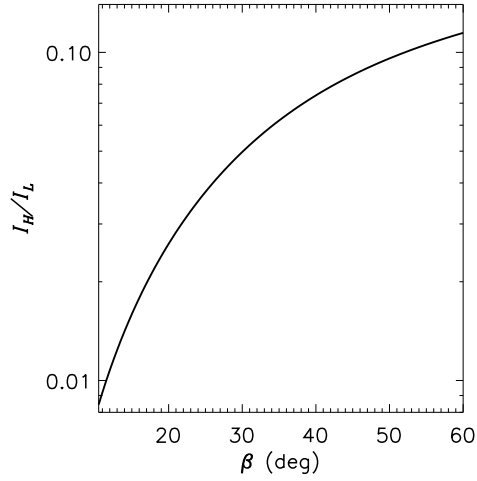}

   \caption{The ratio of integral white-light intensity of halo
   CME to that of normal CME as a function of $\beta$ at a projected
   heliocentric distance 3$R_\odot$.}
   \label{Fig6}
   \end{figure}

The change in brightness caused by Thomson-scattering effect is a little
complicated. For comparability, the white-light intensity is calculated
at a projected distance of $3R_\odot$ for both limb and halo events.
Note that the real heliocentric distance of halo CME become $3R_\odot/
\sin\beta$. The frontal loop is assumed to have a radial width of
0.8$R_\odot$ at 3$R_\odot$ and increases by $1/\sin\beta$. The integral
formula for the LOS white-light intensity \citep{bill66} is then
expressed as

\begin{equation}
I= \frac{1}{2}\pi \sigma_0 J_0 R \int ^{Q}
             _{P}n[(1-u)(2C\cos^{-2}\theta -A)+u(2D\cos^{-2}\theta -B)] d\theta,
\end{equation}\label{eq:LebsequeI}
where the density $n$ is equal to $n_3$ for the limb case and to
$n_3(1/\sin\beta)^{-1.85}$ for the halo case, $n_3$ is the plasma
density of the CME front at $r=3R_{\odot}$, $A=\cos\Omega \sin^2\Omega$,
$B=-0.125[1-3\sin^3\Omega-\frac{\cos^2\Omega}{\sin\Omega}(1+3\sin^3\Omega)\ln\frac{1+\sin\Omega}{\cos\Omega}]$,
$C=(4-3\cos\Omega-\cos^3\Omega)/3$, $D=0.125[5+\sin^3\Omega-\frac{\cos^2\Omega}{\sin\Omega}(5-\sin^3\Omega)\ln\frac{1+\sin\Omega}{\cos\Omega}]$, $\sin\Omega=R_{\odot}/r$, $u=0.6$ is the limb darkening coefficient, $\sigma_0$
is the Thomson-scattering cross section, $J_0$ the photospheric
radiation at the solar surface, and $\theta$ the angular distance
between the plane of the sky and the line connecting the Sun center with
any point along $PQ$. Besides, we assume that the plasma density is
uniform in the frontal loop for simplicity. For different angular
half-width $\beta$, the ratio of the integral intensities of the halo
and limb CMEs, $I_H/I_L$, is plotted in Fig. \ref{Fig6}. It is seen that
for a limb CME, the white-light intensity is reduced by 1-2 orders of
magnitude as it is observed as a halo event. From the figure, we can
also infer that the wider a halo CME really extends, the higher
likelihood it would be detected. Such a result is consistent with
\citet{fain06}, who reported that the real angular widths of halo CMEs
with an average value of $>$60$^\circ$ are relatively larger than that
of normal CMEs.

\begin{figure}
   \centering
   \includegraphics[width=13.0cm, angle=0]{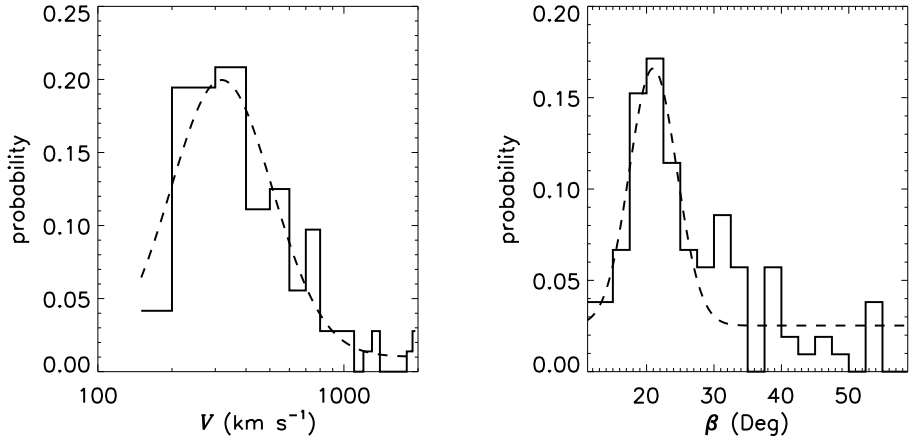}

   \caption{Velocity ({\it Left panel}) and angular half-width ({\it
        Right panel}) distributions of limb CMEs. The histograms are from the
        observational results of \citet{burk04}, while the dashed
        lines are out fitted profiles, a log-normal function for the
        velocity and a Gaussian function for the angular half-width.
   }
   \label{Fig7}
   \end{figure}

\begin{figure}
   \centering
   \includegraphics[width=10.0cm, angle=0]{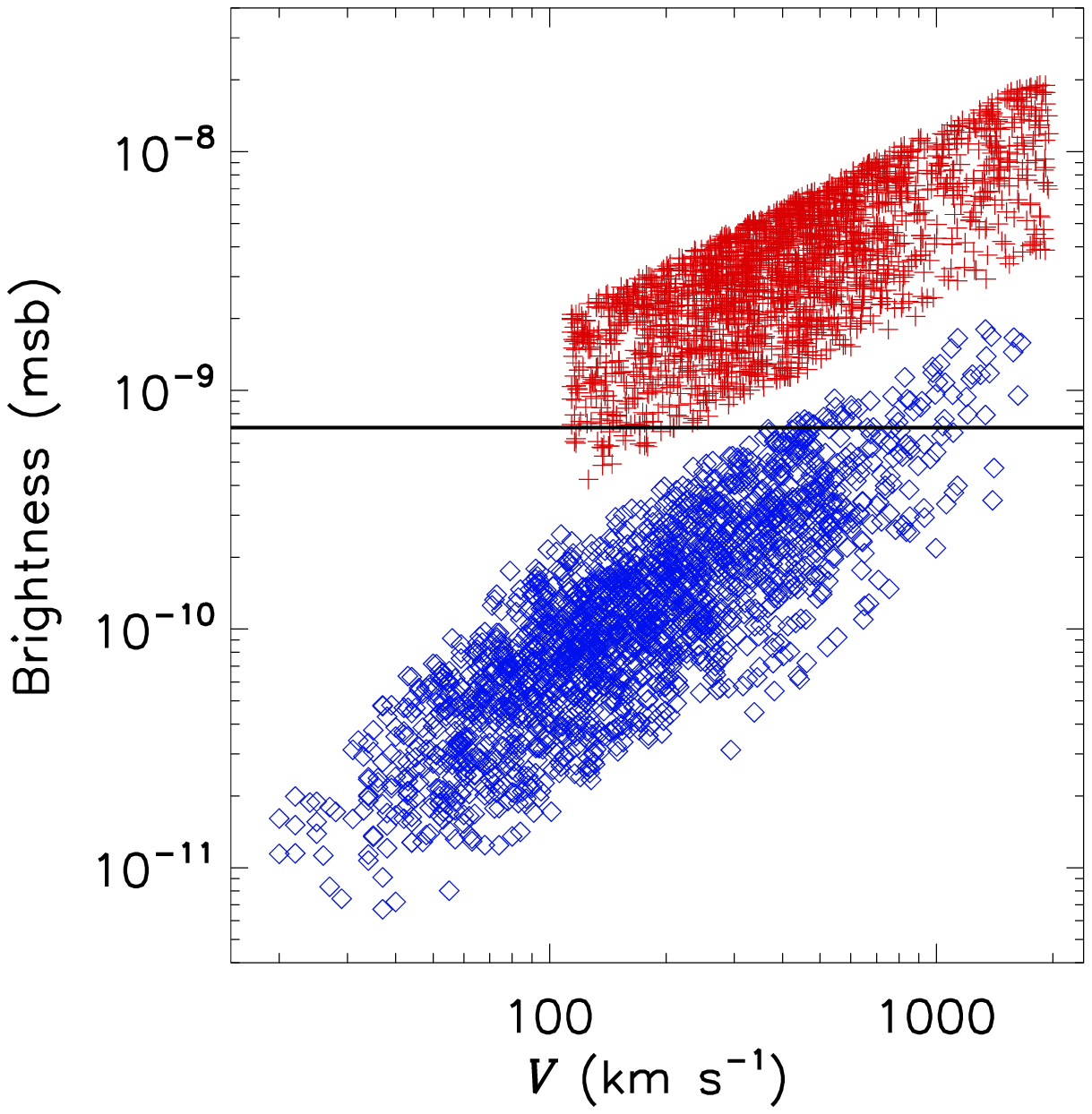}

   \caption{Scatter diagrams of the randomly selected one tenth of the 20\,000
   artificial events. Red crosses and blue diamonds represent limb and halo
   events. The horizontal line marks the 3$\sigma$ level of the background
   solar wind fluctuations at 3$R_\odot$.
   }
   \label{Fig8}
   \end{figure}

In the next step, we perform a Monte Carlo simulation to generate a
sample of 20\,000 artificial limb CMEs that follow the real
velocity and angular width distributions. Generally, both the CME
velocity and the angular width suffer from the projection effects.
In order to eliminate the projection effects, \citet{burk04} studied
the statistical properties of limb CMEs, whose observed velocity and
angular width are almost real. Using their data set, we display the
velocity distribution of the 72 events with velocity $\geq$ 100 km
s$^{-1}$ in the left panel of Fig.~\ref{Fig7}. Similar to
\citet{aoki03}, the profile is fitted with a log-normal function, which
results in $f_V=0.19\exp[-(\ln V-5.76)^2/0.48]+0.01$ as displayed by the
dashed line. The angular width distribution of the 105 events with
$\beta \geq$ 10$^\circ$ is illustrated in the right panel of Fig.
\ref{Fig7}. The histogram is fitted with a Gaussian function $f_\beta=
0.14\exp[-(2\beta-41.95)^2 /7.02]+0.03$, as shown by the dashed line.
The 20\,000 artificial CMEs are assumed to follow these distributions,
with the average velocity being 523 km s$^{-1}$ and the average angular
width being $53.8^\circ$, respectively.

Then, for each velocity interval, data points with different angular
widths are randomly distributed within the two bounding lines shown
in the left panel of Fig.~\ref{Fig4}. The 20\,000 artificial limb CMEs
are scatter-plotted as red crosses in Fig.~\ref{Fig8} in the case of
$k=3.85$. Note that only the randomly selected 10\% of the data points
are shown in order to make the diagram clear. The thick solid horizontal
line marks the 3$\sigma$ level of the white-light noise of the
background solar corona at $3R_\odot$, below which a data point is
considered as unobservable.  Here $\sigma=2.33\times 10^{-10}$ msb is
the standard deviation of the background fluctuations in the LASCO/C2
images at 3$R_\odot$. For each data point, as the event is observed as
a halo CME, its apparent (plane-of-the-sky) velocity decreases by a
factor of $\sin\beta$, while its white-light intensity drops by
$I_H/I_L$ (see Fig.~\ref{Fig6}).  After such corrections, the new data
points of the corresponding halo CMEs are plotted as blue diamonds. It
is found that if the artificial events are observed as halo CMEs, the
white-light intensity of a majority of the sample falls below the
3$\sigma$ level, and therefore the corresponding CMEs are considered to
be ``missed'' by the LASCO coronagraph. The average velocity
($\overline{V}$) of the ``visible'' halo CMEs is calculated to be 922
km s$^{-1}$.

\begin{table}
\centering

\caption{Parameter survey of the average velocity of halo CMEs
($\overline{V}$, in unit of km s$^{-1}$).}
\label{Table-2}

\tabcolsep 6mm
\begin{tabular}{c|ccc}
\hline
    & $h=0.5R_\odot$ & $h=0.8R_\odot$ & $h=1.0R_\odot$ \\
\hline
$k=3.00$ &   894 & 673 & 600 \\
$k=3.85$ & 1 144 & 922 & 826 \\
$k=4.00$ & 1 161 & 954 & 861 \\
\hline
\end{tabular}
\end{table}

There are several assumptions about the properties of the CME frontal
loop in the simulation, e.g., the radial width ($h$), the density
variation with height ($r^{-(k-2)}$), and so on, where the density variation
with height was derived from the 31 well-defined CMEs. To show how the
result is affected by the assumptions, we performed a series of
simulations with different $h$ and $k$. The corresponding $\overline{V}$
is displayed in Table~\ref{Table-2}, with the unit of km s$^{-1}$.

\section{Projection effect} \label{proj}

As one of the puzzles in CME research, it is recognized that halo CMEs
are much faster than normal events, although in principle, the
difference between halo and limb events is only the direction of
propagation. Here we discuss whether the simple projection effect is
related to the high velocity puzzle of halo CMEs.

\begin{figure}
   \centering
   \includegraphics[width=10.0cm, angle=0]{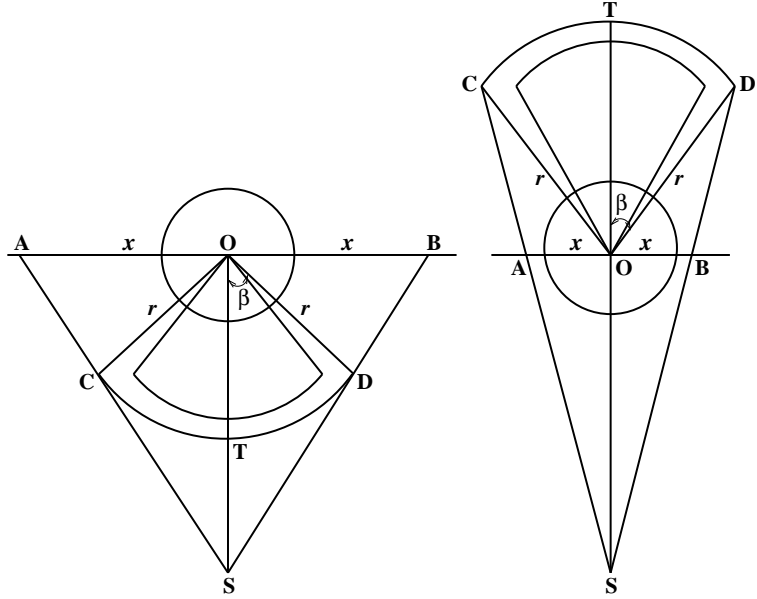}

   \caption{The observing geometry of halo CMEs in the front-side
   ({\it left panel}) and back-side ({\it right panel}) cases. $AB$ signifies
the plane of the sky, and $S$ the observing spacecraft (SOHO).
   }
   \label{Fig9}
   \end{figure}

\begin{figure}
   \centering
   \includegraphics[width=12.0cm, angle=0]{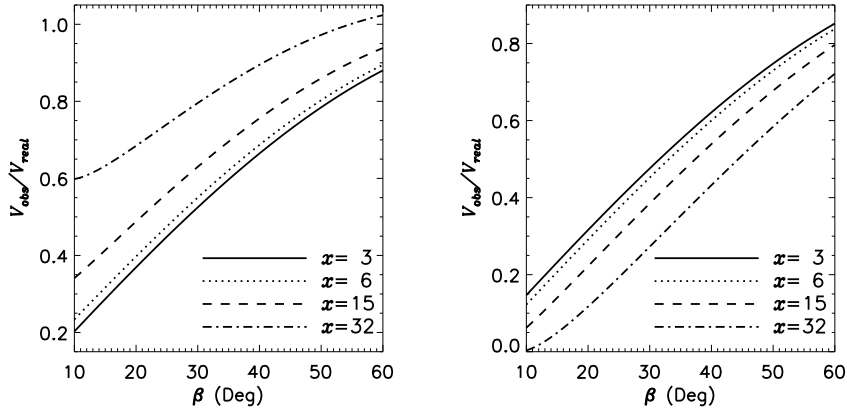}

   \caption{The relationship between $V_\mathrm{obs}/V_\mathrm{real}$
            and $\beta$ for fore-side ({\it Left panel}) and back-side
            ({\it Right panel}) halo CMEs. Different line styles
            (solid, dotted, dashed, and dash-dotted) represent the relationship
            at different projected heliocentric distances (3, 6, 15, and 32$R_\odot$).
   }
   \label{Fig10}
   \end{figure}

Since halo CMEs propagate either toward or away from the Earth as
illustrated in Fig.~\ref{Fig9}, the leading edge (e.g., point $D$)
is projected to the plane of the sky (i.e., point $B$) to calculate
the propagation velocity of the CME. The left/right panel of
Fig.~\ref{Fig9} depicts the propagation geometry for a
front-side/back-side CME with real angular half-width $\beta$. The
real heliocentric distance of the leading loop, $r$, is equal to the
length of $OD$, while the apparent heliocentric distance, $x$, is
equal to the length of $OB$. The SOHO spacecraft, which is located
at $L_1$ with a heliocentric distance of $L= 212R_\odot$, is
marked with S. According to the geometrical relation, the ratio
between the apparent velocity ($V_\mathrm{obs}$) and the real
velocity ($V_\mathrm{real}$) as a function of $\beta$ and $x$ is
expressed as

\begin{equation}
V_\mathrm{obs}/V_\mathrm{real}={\frac{dx}{dt}}/{\frac{dr}{dt}}=\cases{
[1+x/(L\tan\beta)]^2\sin\beta, & front-side, \cr
[1-x/(L\tan\beta)]^2\sin\beta, & back-side.\cr}
\label{eq:LebsequeII}
\end{equation}

The relationship between the velocity ratio and $\beta$ in both
cases is displayed in Fig.~\ref{Fig10} (left panel for the
front-side events and right panel for the back-side), where
different lines correspond to the relation at different apparent
heliocentric distances within LASCO FOV. It is seen that for both
types of events, the apparent velocity is always smaller than the
real one within the LASCO FOV. For the front-side halo CMEs, the
velocity-reducing factor increases from 0.2 to 0.88 as $\beta$
increases from 10$^\circ$ to 60$^\circ$ when the projected
heliocentric distance is 3$R_\odot$. At a larger distance, the
reducing factor increases. For the back-side events, the reducing
factor is slightly smaller than that in the front-side case at
3$R_\odot$. However, the factor decreases with distance as indicated
in the right panel.

The above theoretical analysis reveals that the projection effect
results in a smaller apparent velocity of halo CMEs. Therefore, the
projection effect cannot resolve the high velocity puzzle of halo
CMEs. Such a result is consistent with \citet{Mich03} and
\citet{howa08}, who found that the CME velocity becomes higher after
correcting the projection effect. Combined radio, in situ, and
white-light observations of CME/shocks have also revealed that the
plane-of-the-sky speeds for fast halo CMEs are always less than or equal
to the initial radial speeds \citep{Reiner07}.

\section{Summary} \label{sum}

In this paper we first analyzed the relation between the white-light
intensity and the propagation velocity of 31 limb CMEs observed clearly
by the SOHO/LASCO coronagraph. It is confirmed that slower CMEs tend to
be weaker in the white-light brightness, meaning that they carry less
plasma. We also studied the normalized intensity evolution of the CME
leading edge along with the radial distance $r$ and fitted it with a
power-law function, $I_{n}\sim r^{-k}$. It is found that the power index
$k$ is averaged around 3.85. Considering that the incident emission from
the photosphere decreases as $r^{-2}$, it means that the plasma density
of a CME frontal loop, $n$, decreases with $r$ roughly as $r^{-1.85}$.
Although the cross-section of Thomson scattering increases as a limb CME
is observed as a halo CME, the abrupt decreases of the incident light
from the photosphere and $n$ with $r$ lead to that the white-light
intensity of halo CMEs would be very weak since they should travel a
longer distance to be observed by coronagraphs.

As a further quantitative investigation to resolve the high velocity
puzzle for halo CMEs, we performed a Monte Carlo
simulation of 20\,000 artificial CMEs with velocity and angular width
distributions derived from limb CMEs. The Thomson-scattering intensity
is calculated for both limb and halo events. The simulation indicates
that if the limb CMEs with angular width between 20$^\circ$ and
120$^\circ$ are observed as full halo events propagating along the
Sun-Earth line, a majority of the slower events become so weak that
their intensity is comparable to the fluctuation of the background
solar wind, implying that they would fail to be identified by
coronagraphs. The average velocity of the ``detectable'' halo CMEs is
$\sim$922 km s$^{-1}$, which agrees perfectly with previous statistical
result \citep{Mich03}. We believe that the missing of many slow halo
CMEs can well explain the high velocity puzzle of halo CMEs. As a proof
of statement that some CMEs may be neglected even by the state-of-art
coronagraphs, here we mention the 2007 December 7 event, which was
observed by STEREO (Solar TErrestrial RElations Observatory) A satellite
\citep{kai05}, but completely missed by STEREO B since the event was
more face-on to STEREO B \citep{Ma09}.  Statistical work using
multi-directional data is of help to clarify how many faint halo CMEs
are missed and testify our conclusion.

It should also be emphasized that the halo CMEs in the simulation are
full halo ones that originate at the disk center and propagate along
the Sun-earth line. For the full halo CMEs that propagate aside from the
Sun-earth line and partial halo CMEs, the white-light intensity decrease
would not be so significant as in Fig.~\ref{Fig6}. Thus, less portion
of such events would be missed. In order to quantitatively compare the
Monte Carlo simulation with observations, these cases would also be
considered in a future work.

\normalem
\begin{acknowledgements}
The authors thank J. Zhang and Y. H. Tang for instructive discussions
and suggestion throughout the work. We are also grateful to the
anonymous referee for suggestions. This research is supported by the
Chinese foundations GYHY200706013, 2006CB806302, NSFC (10403003,
10933003, and 10673004). SOHO is a project of international cooperation
between ESA and NASA.
\end{acknowledgements}

\label{lastpage}

\end{document}